\documentclass[conference]{IEEEtran}
\IEEEoverridecommandlockouts
\usepackage{cite}
\usepackage{amsmath,amssymb,amsfonts}
\usepackage{algorithmic}
\usepackage{graphicx}
\usepackage{textcomp}
\usepackage{xcolor}
\usepackage{amsthm}
\newtheorem{definition}{Definition}
\usepackage{tikz}
\usepackage{amsthm}
\usepackage{amssymb}
\newtheorem{theorem}{Theorem}[section]

\newtheorem{proposition}[theorem]{Proposition}

\usepackage{caption}
\usepackage{bbm}
\def\BibTeX{{\rm B\kern-.05em{\sc i\kern-.025em b}\kern-.08em
    T\kern-.1667em\lower.7ex\hbox{E}\kern-.125emX}}

\graphicspath{{./figs/}}
\newcommand\inputpgf[2]{{
    \let\pgfimageWithoutPath\pgfimage
    \renewcommand{\pgfimage}[2][]{\pgfimageWithoutPath[##1]{#1/##2}}
    \input{#1/#2}
}}

\begin{document}

\title{A Propagation-model Empowered Solution for Blind-Calibration of Sensors\\}

\author{\IEEEauthorblockN{Amit Kumar Mishra}
\IEEEauthorblockA{\textit{Department of Electrical Engineering} \\
\textit{University of Cape Town}\\
Cape Town, South Africa  \\
akmishra@ieee.org}
}

\maketitle

\begin{abstract}
Calibration of sensors is a major challenge especially in inexpensive sensors and sensors installed in inaccessible locations. 
The feasibility of calibrating sensors without the need for a standard sensor is called blind calibration. 
There is very little work in the open literature on totally blind calibration. 
In this work we model the sensing process as a combination of two processes, viz. propagation of the event through the environment to the sensor and measurement process in the sensor. 
Based on this, we propose a unique method for calibration in two flavours, viz semi-blind and completely-blind calibration. 
We show limited results based on simulation showing encouraging results.
\end{abstract}

\begin{IEEEkeywords}
Sensors, Calibration, Blind-calibration, Sensor Network, AI
\end{IEEEkeywords}

\section{Introduction}
Calibration is a major component of any metrological system especially for sensors which are installed in remote places. Without a through investigation and methodology around sensor calibration, usually the data collected from the sensors are not reliable. In one of the very few honest papers in the open literature Bittner etal  \cite{bittner2022performance} discussed how they got their sensors calibrated to a high standard and then those were installed in Malawi. However, they observed how quickly the quality of the data from the sensor network became almost non usable. 
This is a major pain point in the current day of ubiquitous sensing.

One can find a summary review of in situ calibration methods in \cite{delaine_19situ}. 
Calibration efforts for individual sensor types are extensive. 
Most of these processes need a reference sensor or some ground-truths. For example in a work on low cost air pollution sensors in Norway \cite{veiga2021low}, the researchers used reference based calibration.  
Running a reference-based calibration for remote sensors is a costly task. Also, it does not scale up. I.e. when the number of sensors is in hundreds the task becomes impossible to be carried out on a regular basis.

One of the solutions to this challenge has been to treat the battery of sensors as a single system. This system can, then, be calibrated as a whole  rather than focusing on individual sensors in this network. In their pioneering work, Whitehouse etal \cite{whitehouse2002calibration} used a physics based model. 
The data from the sensor network is expected to fit the model as closely as possible. Hence, the individual sensor-calibration parameters are fine-tuned to force this fit.
Many following works have used this approach and modified it as well. Though efficient, this approach is not fully blind. 

The second important piece of work in the domain of blind calibration was presented by Balzano and Nowak \cite{balzano2007blind} in their work on blind sesnor network calibration methodology. Their assumption of the existence of spatial oversampling gave an elegant solution which has been leveraged upon by many other works since then. For example, in a recent work \cite{wang2017deep} machine learning has been used to learn the sub-space projection part of Balzano's method. 
Though elegant, this method does not work when the number of sensors is not too many. Unfortunately, in most real life cases, the number of sensors available is usually limited. 
However, when it comes to the methodologies of blind calibration processes for a single sensor or a limited number of sensors (negating the oversampling assumption) there is not many reports in the open literature. 

In this work we model the sensing process as a combination of two processes, viz. propagation of the event through the environment to the sensor and measurement process in the sensor. 
Based on this, we propose a unique method for calibration in two flavours, viz semi-blind and completely-blind calibration.  
It can be noted here that this set of methods is suitable for both a sensor network as well as a single sensor. 

Following are the main claims about the invention. 
 The invention can be used to regularly calibrate sensors in inaccessible locations.  
  The invention proposes two methods, one which requires some intervention by the user during the calibration process and in the other no intervention by the user is needed. 
  The invention has two parts. In one part it uses conventional numerical methods and in the other it uses modern machine learning methods. 
  The invention can be used for either single sensor or a network of sensors.

The rest of the paper is organized as follows. In the next section we shall present the model we shall use in our work. 
Section III will propose the solution followed by the two flavours of it (autonomous and semi-autonomous). 
The paper shall end with some concluding remarks.

\section{Model of the Measurement Process}
For the sake of generality, let us assume that we are dealing with a single sensor. 
The treatment can be extended to multiple sensors trivially.  

The intention of any measurement system is to measure an event, $e$, which has created a space-time filed. 
The sensor is used to measure a measurand (true variable of interest), $x$, at a given location and time.  
Output from the sensor is $y$. 
The event, $e$, has been modulated by the spatio-temoral response of the environment to create the measurand field from which $x$ is measured at a given location and time. 

\begin{definition}
If we represent time with $t$ and spatial vector by $\Vec{s}$, then we can represent the spatio-temporal response of the environment by $h(t,\Vec{s})$. 
If we assume the process to be linear then the measurand field, $x(t,\Vec{s})$, can be represented as: 
\begin{equation}
    x(t,\Vec{s}) = h(t,\Vec{s}) \ast e(t,\Vec{s}),
\end{equation}
where $\ast$ represents the convolution operation.
\end{definition}
 
\begin{definition}
The Observation Process \cite{cox2008probabilistic} (process that outputs $y$ from $x$) can be modelled by the function $f$ whereas the Restituition Process (of getting the true measurand back from the sensor readings) is modelled by $f^{-1}$. 
\end{definition}

Hence, the signal measured by a given sensor at a given time $t$ would be 
\begin{equation}
    y(t,\Vec{s}) = f( x(t,\Vec{s})) + n(t) 
    = f(h(t,\Vec{s}) \ast e(t,\Vec{s})) + n(t),
\end{equation}
where $n(t)$ is the measurement noise that can not be modelled by the processes. We have assumed it to be additive white Gaussian noise (AWGN).  Hence, we have not made it depend on the location $\Vec{s}$.
Similarly, 
\begin{equation}
    \Tilde{x}(t,\Vec{s}) = f^{-1}(y(t,\Vec{s})).
\end{equation}
 The measurement expression in Equation 1 can be rewritten as:
\begin{equation}
    y(t,\Vec{s}) =  f(h(t,\Vec{s}) \ast e(t,\Vec{s})) + n(t) = f(h(t,\Vec{s})) \ast f(e(t,\Vec{s})) + n(t). 
\label{measurement_equation}
\end{equation}
To make the analysis simpler we can assume that we only have one sensor and drop the location parameter. 
\begin{equation}
    y(t) =  f(h(t) \ast e(t)) + n(t) = f(h(t)) \ast f(e(t)) + n(t). 
\end{equation}

\begin{figure}
    \centering
    \includegraphics[scale=0.3]{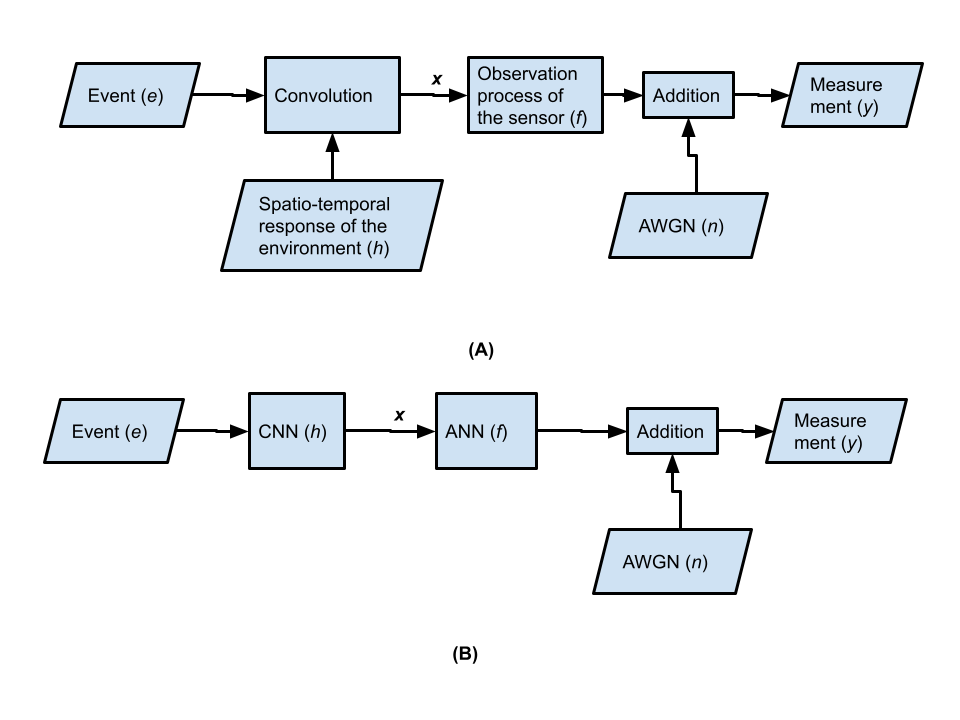}
    \caption{Measurement process chain in a sensor system. (A) shows the chain using the conventional signal processing blocks (as described in Equation ~\ref{measurement_equation}. (B) shows the chain by replacing some of the blocks with machine learning based processing blocks. }
    \label{measure}
\end{figure}

The signal flow graph of the above expressions is shown in Figure~\ref{measure}-A. 
 The signal processing blocks can also be replaced by their equivalent machine learning blocks (as shown in Figure ~\ref{measure}-B). 
 It can be noted here that the use of artificial neural network (ANN) for observation process (and hence in reference-based calibration) is an active field of research\cite{yamamoto2017machine, concas2021low}.  



\section{Proposed Design}
Let us list all the variables and functions that we are working with in this model as discussed in the previous section. 
\begin{itemize}
    \item $e$: The event causing the measurand.
    \item $h$: The spatio-temporal response of the environment.
    \item $x$: The measurand at the sensor.
    \item $f$: The observation process of the sensor. 
    \item $y$: The measured value of the sensor.
\end{itemize}

We can divide the life of the sensor in the field into two stages. 
\begin{enumerate}
    \item Reliable Calibration (RC) Stage: This is when the sensor has been installed in the filed recently and the calibration model is reliable. In this case we know two of the variables/functions from the above list, viz. $y$ and $f$. 
    \item Unreliable Calibration (UC) Stage: This is when the sensor has been operating in the filed for longer time than what it would take to create errors and drifts in the calibration function. In this case we only know $y$ reliably.
\end{enumerate}

The purpose of calibration is to make sure that we can estimate the changes to the sensor response function $f$ that happens over time or due to changes in the operational conditions.

\begin{proposition}
{\bf Using $h$ for Calibration:} The response of the environment, $h$, does not depend on the aging of the sensors. Hence, $h$ can be used as a process-invariant. If $h$ can be estimated in the RC stage (where $f$ is known correctly) then in the UC stage, the known $h$ can be used to estimate $f$.
\end{proposition}

We shall propose two novel strategies to do this. 
\begin{itemize}
    \item {\bf Semi-blind Calibration:} In this, we shall use controlled perturbation in $e$. However, we shall propose ways so that this can be done by a non-expert and the process would be robust to unpredictable changes in the environment.
    \item {\bf Blind Calibration:} In the cases where the above semi-blind calibration process is impossible to carry out, we shall model the measurement process as a two stage autoencoder network \cite{goodfellow2016deep} and propose a fully blind calibration process. 
\end{itemize}

\subsection{Semi-blind Calibration}
We assume the existence of a way to create a controlled profile of perturbation in the measurand field. This is usually possible. For example, if we are considering particulate material (PM) sensors, we can create a unit function based perturbation of PM in the vicinity. 

The calibration in this case shall be carried out following two steps. 
\\
{\bf Step 1:} Let us refer to the measurement system Equation  ~\ref{measurement_equation}. 
We are creating a known perturbation, $e$, in the measurand field and in  the RC stage we know $f$ correctly. Hence, in the RC stage we know all the terms except $h$ and the AWGN part. 
From this we can use either numerical methods (following the system diagram as given in Figure ~\ref{measure}-A) or machine learning methods (following the system diagram as given in Figure ~\ref{measure}-B) to estimate $h$. 
\\
{\bf Step 2:} In the UC stage, we do not know $f$. However, we know $h$ from the previous step. We can use numerical methods to estimate $h$. From this we can know the deviation in $h$, $\Delta h$. 
This can be used either through numerical methods (following the system diagram as given in Figure ~\ref{measure}-A) or machine learning methods (following the system diagram as given in Figure ~\ref{measure}-B) to estimate $\Delta f$, which then can be used to correct $f$.  
This completes the calibration process. 

In the above steps, we have assumed that $h$ is invariant. This, in fact, is not strictly the case. Hence, using $\Delta h$ is not the ideal. 

\begin{proposition}
Compared to the exact value of $h$, the shape of $h$ is less dependant on the effects of variations in the environment. 
\end{proposition}

Following the above proposition, we shall use the change in the shape of $h$ (rather than the exact value of $h$) to update $f$ in the calibration steps proposed above. 

\subsection{Blind Calibration}
\begin{figure}
    \centering
    \includegraphics[scale=0.3]{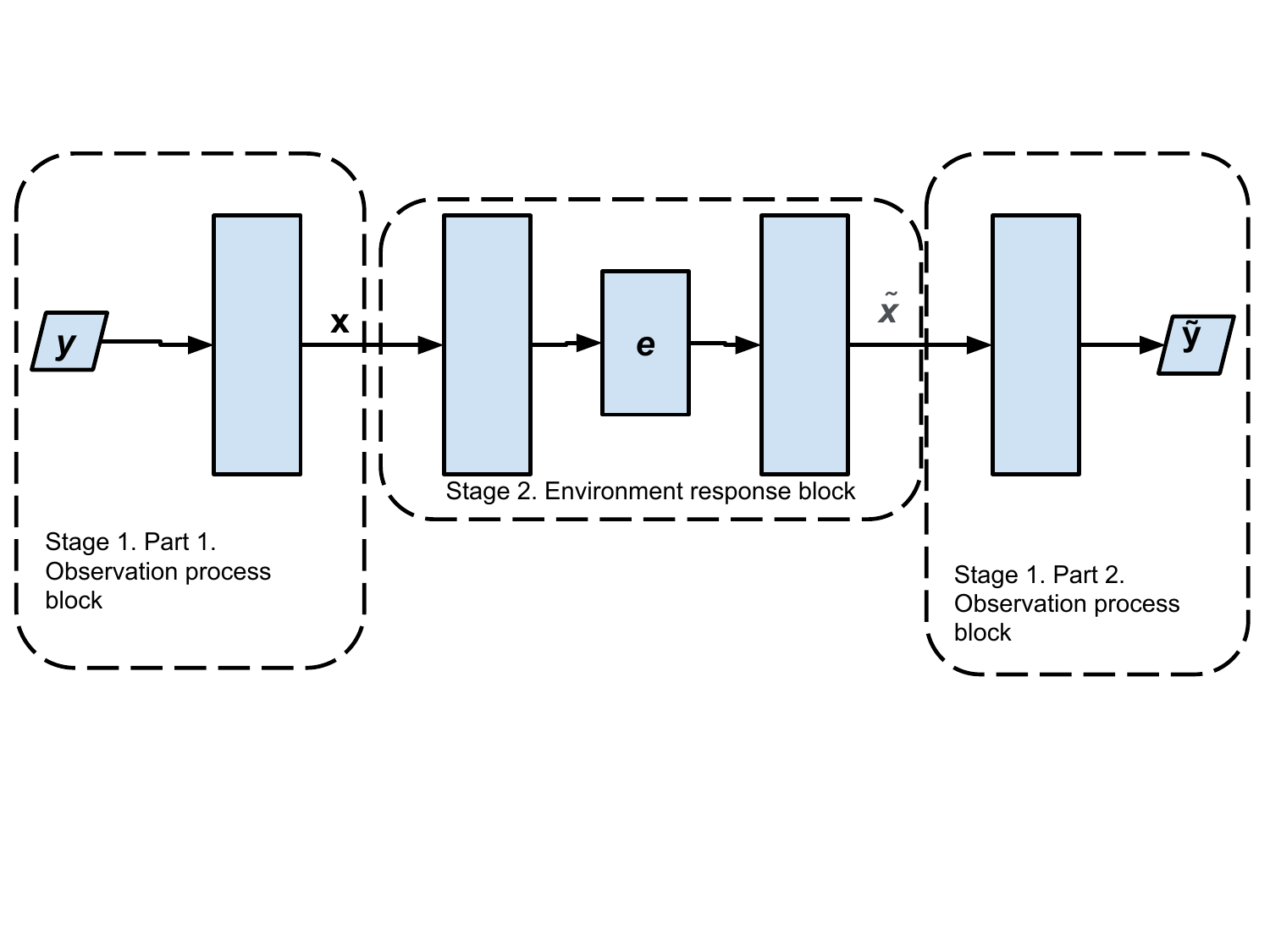}
    \caption{A two stage autoencoder modelling the measurement system. The measured data from the sensor, $y$, shall be used to train this model. 
It has two stages (implemented by two blocks of convolutional neural networks (CNNs)). The first stage is used to model the data dependency from $y$, the measured data from the sensor, to the measurand field, $x$. 
The second stage is used to model the  data dependency from $x$, the the measurand field, to the event field, $e$.}
    \label{auto}
\end{figure}

Without the existence of a known perturbation $e$ it is impossible to run the above calibration method. We shall have more unknowns than known data-sets. 
As we do not have enough information about the measurement process (Equation ~\ref{measurement_equation}), we can rely on the structure of the signal flow graph.  This can, then, be used to design an autoencoder network with two stages as shown in Figure~\ref{auto}. 
The measured data from the sensor, $y$, shall be used to train this model. 
It has two stages (implemented by two blocks of convolutional neural networks (CNNs)). The first stage is used to model the data dependency from $y$, the measured data from the sensor, to the measurand field, $x$. 
The second stage is used to model the  data dependency from $x$, the the measurand field, to the event field, $e$. 

\begin{proposition}
{\bf Autoencoder-based Calibration:} The environment response block models the relatively invariant function $h$. Hence, by using a two-stage autoencoder, the calibration function is captured by the observation process block.
\end{proposition}

The calibration in this case shall be carried out following two steps. 
\\
{\bf Step 1:} In the RC stage, the network is trained using both $x$ and $y$ (because we know $f$ correctly). In this step, both the observation process and environment response blocks get trained.
\\
{\bf Step 2:} In the UC stage, the network is trained again. In this case we only know $y$, the sensor readings. However, in this step, the environment response block is not changed. Hence, only the observation process block gets updated. 
After this step, the updated observation process is used as the updated calibration process. 

\subsection{Sensor Networks}
The above two methods can be used either for single sensor or for a network of sensors. In fact, by having a network of sensors, we get more information about the shape of $h$. Hence, the calibration process will be more accurate. 

\section{Simulation based Validation}
Validating Propositions III.1 and III.2 would need experimental procedures. These will be done in the future work. 
However, Proposition III.3 can be performed by simulation. 
Being the derived proposition, validating this would also offer limited proof for the potential working of the original proposition, Proposition III.1. 

\subsection{Experimental Setup}
We list the way we have simulated and modelled the different signals and processes of the environment (as shown in Figure \ref{measure}.

\begin{itemize}
    \item The event to be measured, $e$, is modelled as an AWGN. 
    \item The environmental response is modelled by a 100-tap band-pass filter, $h$. 
    \item The calibration function is modelled by a polynomial of degree three, $$f(x) = x + k_1 x^2 + k_2 x^3.$$ 
    \item The signal $y$ is fed in batched of 128 data to the autoencoder. 
    \item {\bf The autoencoder} has four fully connected layers. As our aim is not to compress or find a reduced space representation, we have kept the dimensions the same for each layer. The error function is a function of the error between $x$ and $\Tilde{x}$ as well as between $y$ and $\Tilde{y}$, i.e.  $$\alpha L(x,\Tilde{x}) + \beta L(y, \Tilde{y}).$$ $L()$ is the loss function and in this case we have chosen mean square error to be our loss function. The parameters $\alpha$ and $\beta$ are used to set the relative importance between the two components of the error function. As of now, we have chosen them both to be equal to one. In the future, we shall investigate the effect of differential weighting on the calibration process.
\end{itemize}
It can be noted here that the coefficients $k_1$ and $k_2$ represent the sensor calibration process and are assumed to drift with aging. 
For a well calibrated sensor, these two coefficients will be very close to zero and will slowly increase with aging. 

\begin{figure}
    \centering
    \includegraphics[scale=0.5]{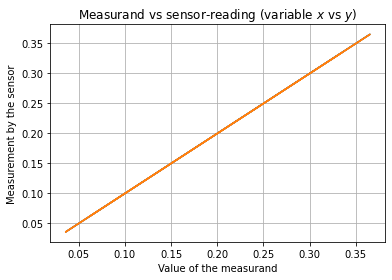}
    \caption{Measurand and measured values, $x$ and $y$, with well calibrated sensor. The two lines overlap on each other and are indistinguishable.}
    \label{cal1}
\end{figure}

\begin{figure}
    \centering
    \includegraphics[scale=0.5]{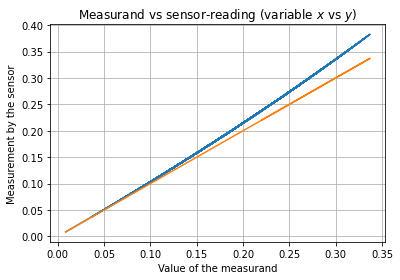}
    \caption{Measurand and measured values, $x$ and $y$, when the measurement process has drifted. Hence, the calibration process no longer is able to estimate the true measurand value. This is visible from the divergence between the two plots. }
    \label{cal2}
\end{figure}

In our experiments, we start with $k_1 = 0.001$ and $k_2 = 0.0001$. This shows a calibrated sensor. Figure \ref{cal1} shows the plots of these two signals. The two plots are indistinguishable. 
For the uncalibrated situation we use $k_1 = 0.3$ and $k_2 = 0.3$. Figure \ref{cal2} shows the plots of these two signals.


In the first set of training, we used the calibrated data to train the autoencoder. In the second stage, we use the higher values of $k$ as mentioned above. However, this time, we do not train the whole network. 
Rather, we train the network between $y$ and $x$, and $\Tilde{x}$ and $\Tilde{y}$.

\begin{figure}
    \centering
    \includegraphics[scale=0.5]{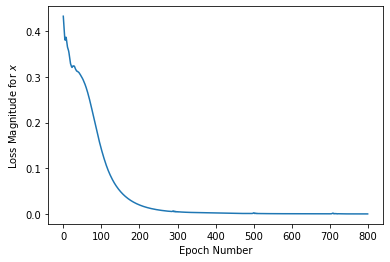}
    \caption{The decrease in loss function for $x$ (the measurand) with training of the autoencoder for the calibrated case. It can be noted here that this is two-stage autoencoder and has two values it needs to learn to predict, viz. $x$ and $y$.}
    \label{lossx}
\end{figure}

\begin{figure}
    \centering
    \includegraphics[scale=0.5]{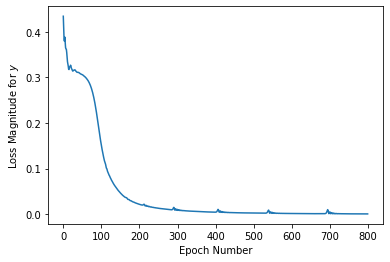}
    \caption{The decrease in loss function for $y$ (the data measured by the sensor) with training of the autoencoder for the calibrated case. It can be noted here that this is two-stage autoencoder and has two values it needs to learn to predict, viz. $x$ and $y$.}
    \label{lossy}
\end{figure}

\begin{figure}
    \centering
    \includegraphics[scale=0.5]{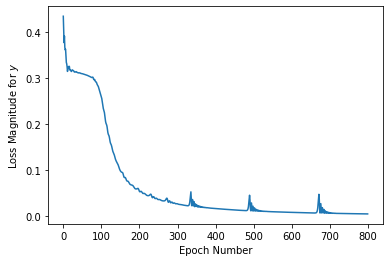}
    \caption{The decrease in loss function for $y$ (the data measured by the sensor) with training of the autoencoder for the uncalibrated case. Even though the loss function does not decrease as smoothly as the previous cases, it does converge and shows that the training process gets completed successfully. }
    \label{lossycal}
\end{figure}

The autoencoder is trained with the initial condition of calibrated sensor ($k_1 = 0.001$ and $k_2 = 0.0001$). 
Figure \ref{lossx} shows the training process of the autoencoder in terms of learning the value of $x$ the true measurand. \\
Figure \ref{lossy} shows the training process of the autoencoder in terms of learning the value of $y$ the value measured by the sensor. 
These figures show a successful training of the autoencoder. \\
Later, we use the data with the uncalibrated case ($k_1 = 0.3$ and $k_2 = 0.3$). In this phase of training only the connections between $x$ and $y$ and $\Tilde{x}$ and $\Tilde{y}$ are updated. Figure \ref{lossycal} shows the training process in terms of loss with respect to the number of epochs. It can be noted that even though the loss function does not decrease as smoothly as the previous cases, it does converge and shows that the training process gets completed successfully.  


\begin{figure}
    \centering
    \includegraphics[scale=0.5]{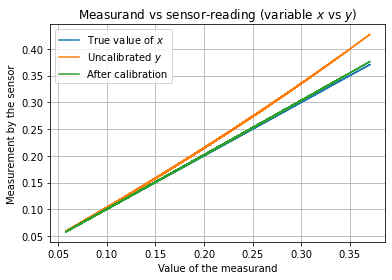}
    \caption{Measurand and measured values, $x$ and $y$, with well calibrated sensor. The two lines are indistinguishable}
    \label{calprocess}
\end{figure}

Next, we use the trained network (between $y$ and $x$) to calibrate the sensor readings. These data are plotted (along with the data from calibrated and uncalibrated cases) in Figure \ref{calprocess}. It can be seen that the error caused by the drifts in $k_1$ and $k_2$ has been corrected significantly. This validates the use of our proposed novel calibration method. 

\section{Conclusion and Future Work}
In this work, we have presented a new method\footnote{The invention has been filed as a patent in the UK patent office with the title, “Method and System of Calibration of a Sensor or a network of Sensors”, and  application number 2215800.0 (filed on 25-Oct-2022).} for blind and semi-blind calibration of sensor(s) by using machine learning paradigms. 
We have proposed three methods, two of which are semi-blind and the third one is blind. 
We have validated the semi-blind calibration proposition with limited simulation results. 
To our limited knowledge, this is the first time a completely blind method of calibration has been proposed in the open literature. Of course, the method shall not work for indefinite period. However, this will increase the inter-CalVal-routine time. In other words, a full scale calibration validation exercise would be required less often. This is a major benefit in industrial setups. 
The other use of the proposed methods can be in the field of inexpensive sensor development for the measurement in difficult to access regions or in places where there is no available sensing currently. For example, in our group we are developing inexpensive sensor modules to measure various types of air pollutants in African cities. In another project we are developing sensor systems to measure wave parameters in the Southern Ocean. In these kinds of applications, sensors are not, usually, calibrated once they have been put in the field. For such usages, our methods can prove as a major benefit. Because, currently there are no methods to re-calibrate the sensors. In such situations, our method can help the sensors to measure reliable data for longer duration of time. 

\section{Acknowledgement}
This work has been supported with funding from Sentech Soc Ltd, South Africa.

\bibliographystyle{IEEEtran}
\bibliography{asin_bib}

\end{document}